\documentclass[oneside, 12pt]{article}

\usepackage[margin=3cm]{geometry}

\usepackage[utf8]{inputenc}
\usepackage[english]{babel}

\usepackage{amsthm}   
\usepackage{amsmath}  
\usepackage{amssymb}  
\usepackage{bm} 
\usepackage{verbatim} 
\usepackage[inline]{enumitem} 
\usepackage{cancel}   
\usepackage{multicol}
\usepackage[dvipsnames]{xcolor} 
\usepackage{multirow} 
\usepackage{booktabs} 
\usepackage[font=footnotesize]{caption} 
\usepackage{tikz-cd}
\usepackage{xurl}
\usepackage{hyperref}
\hypersetup{
  colorlinks=true,
  citecolor=darkgray,
  linkcolor=darkgray,
  urlcolor=darkgray,
  breaklinks=true,
  bookmarksnumbered=true,
  bookmarksopen=true,
  pdftitle={Towards a Coq formalization of a quantified modal logic},
  pdfauthor={Ana de Almeida Borges},
  pdfsubject={},
  pdfcreator={Ana de Almeida Borges},
  pdfkeywords={}
}
\usepackage[numbers]{natbib}   

\theoremstyle{definition}
\newtheorem{theorem}{Theorem}[section]
\newtheorem{definition}[theorem]{Definition}
\newtheorem{lemma}[theorem]{Lemma}

\newcommand{\ana}[1]{}
\newcommand{\mireia}[1]{}

\newcounter{dummy}
\makeatletter
\newcommand\myitem[1][]{\item\refstepcounter{dummy}\def\@currentlabel{#1}}
\makeatother

\newcommand{\Forall}[1]{
  \forall #1 \,
}

\usepackage{xifthen}

\newcommand{\coqdocbasebaseurl}{https://ana-borges.gitlab.io/QRC1-Coq/v0.1.0}

\newcommand{\coqdocbaseurl}{\coqdocbasebaseurl /QRC1.}
\newcommand{\urlhash}{\#}
\newcommand{\coqdocurl}[2]{\coqdocbaseurl #1.html\urlhash #2}
\newcommand{\nolinkcoqident}[1]{\texttt{\detokenize{#1}}}
\makeatletter
\newcommand{\coqident}{\begingroup\@makeother\#\@coqident}
\newcommand{\@coqident}[3][]{%
  \ifthenelse{\isempty{#2}}%
  {\nolinkcoqident{#3}}%
  {\ifthenelse{\isempty{#1}}%
  {\href{\coqdocurl{#2}{#3}}{\nolinkcoqident{#3}}}%
  {\href{\coqdocurl{#2}{#3}}{\nolinkcoqident{#1}}}}%
\endgroup}
\newcommand{\coqfile}[2]{%
  \ifthenelse{\isempty{#1}}%
    {\href{\coqdocbaseurl #2.html}{\nolinkcoqident{#2.v}}}%
    {\href{\coqdocbaseurl #1.#2.html}{\nolinkcoqident{#2.v}}}}

\makeatother

\newcommand{\coqstdliblink}{https://coq.inria.fr/distrib/V8.14.1/stdlib}
\newcommand{\mathcomplink}{https://math-comp.github.io/htmldoc}

\newcommand{\eqType}{%
  \href{\mathcomplink/mathcomp.ssreflect.eqtype.html}{%
    \texttt{eqType}%
  }%
}

\newcommand{\choiceType}{%
  \href{\mathcomplink/mathcomp.ssreflect.choice.html}{%
    \texttt{choiceType}%
  }%
}

\newcommand{\countType}{%
  \href{\mathcomplink/mathcomp.ssreflect.choice.html}{%
    \texttt{countType}%
  }%
}

\newcommand{\la}{\langle}
\newcommand{\ra}{\rangle}

\newcommand{\RC}{\ensuremath{{\sf RC}}}

\newcommand{\GL}{\ensuremath{{\sf GL}}}

\newcommand{\PA}{\ensuremath{{\sf PA}}}

\newcommand{\QRC}{\mathsf{QRC_1}}

\newcommand{\M}{\mathcal{M}}
\newcommand{\F}{\mathcal{F}}

\newcommand{\fv}{\normalfont{\text{fv}}} 
\newcommand{\subst}[2]{[#1 {\leftarrow} #2]} 
\newcommand{\rsubst}[3]{[#1, #2 {\leftarrow} #3]}
\newcommand\rest[1]{|_{#1}}

\newcommand{\xaltern}[1]{\sim_{#1}} 

\title{Towards a Coq formalization of a quantified modal logic}
\author{Ana de Almeida Borges\thanks{ORCID: 0000-0001-5152-198X}}
\date{}

\begin{document}

\maketitle

\begin{abstract}
  We present a Coq formalization of the Quantified Reflection Calculus with one modality, or $\QRC$. This is a decidable, strictly positive, and quantified modal logic previously studied for its applications in proof theory. The highlights are a deep embedding of $\QRC$ in the Coq proof assistant, a mechanization of the notion of Kripke model with varying domains and a formalization of the soundness theorem. We focus on the design decisions inherent to the formalization and the insights that led to new and simplified proofs.
\end{abstract}


\section{Introduction}
\label{sec:intro}


The Quantified Reflection Calculus with one modality, denoted by $\QRC$ and introduced in \cite{QRC1}, is a strictly positive quantified modal logic inspired by the unimodal fragment of the Reflection Calculus, $\RC_1$ \cite{Dashkov2012, Beklemishev2012}.
  The quantified strictly positive language consists of a \emph{verum} constant and relation symbols as atomic formulas, with the only available connectives being the conjunction, the diamond, and the universal quantifier. $\QRC$ statements are assertions of the form $\varphi \leadsto \psi$ where $\varphi$ and $\psi$ are in this strictly positive language.

  $\QRC$ was born out of the wish for a nice quantified provability logic for theories of arithmetic such as Peano Arithmetic ($\PA$), even though Vardanyan \cite{Vardanyan1986} showed that this is impossible in general. In fact, the full quantified provability logic of $\PA$ is $\Pi^0_2$-complete, and thus not recursively axiomatizable, let alone decidable.
  However, restricting the language to the strictly positive fragment is a viable solution \cite{Escape}.

  The main results obtained for $\QRC$ and described in \cite{Escape} are soundness with respect to varying domain Kripke models, completeness for finite and constant domain Kripke models, and soundness and completeness with respect to two different (but related) arithmetical interpretations, marking it as a provability logic.

  Here we report on an ongoing formalization \cite{QRC1_Coq_v0.1.0} of part of the work presented in \cite{Escape}. We will sometimes cite \cite{QRC1} as well, since it includes a more detailed, albeit less general, version of some of the same results. The current paper focuses on the formalization of the language and axiomatization of $\QRC$ (Sections~\ref{sec:language} and \ref{sec:QRC1} respectively), as well as of its Kripke semantics (Section~\ref{sec:semantics}) and soundness (Section~\ref{sec:soundness}). The formalization of the Kripke completeness is ongoing and will be described in a future work. The formalization of the arithmetical results has not been tackled yet.

\subsection{Related work}

Quantified modal logic has been extensively studied \cite{Goldblatt2011}, and even formalized. For example, \cite{Basin1998} describes a modular Isabelle formalization of several quantified modal logics, including soundness and completeness theorems for them. On the other hand, \cite{BenzmullerPaleo2015} describes a set of Coq tactics to facilitate showing that a user-defined and possibly quantified modal logic proves a given statement. We have not made use of this library as our main goal was to prove meta-theorems of $\QRC$, for which a deep embedding is more appropriate. There has also been work on a custom proof assistant for quantified modal logic \cite{Libal2018}, as well as an automated theorem prover for normal quantified modal logics \cite{Gleissner2017}.

Furthermore, there are several implementations of propositional modal logics, both in Coq \cite{DoczkalSmolka2011, DoczkalBard2018, Worms_Coq} and in other proof assistants \cite{MaggesiBrogi2022}, as well as presentations of first-order logics \cite{OConnor2005, Forster2021}.

\subsection{External Tools}
\label{sec:tools}

Coq \cite{coq} is a general purpose interactive and formal proof management system. It provides a formal language expressive enough to write theorem statements and their proofs, as well as specifications of algorithms and their implementations. These proofs are verified by the Coq kernel, and are thus correct up to hypothetical (and unexpected) errors in the implementation of the kernel itself \cite{Sozeau2019}.
  Coq has been extensively used to formalize both mathematical theorems \cite{OConnor2005, Gonthier2008, DoczkalSmolka2011, Gonthier2013_OddOrder, Hales2017, Doczkal2018, DoczkalBard2018} and software correctness \cite{CompCert, TimeLibrary}.

The core language is called Calculus of Inductive Constructions, a constructive type theory with support for inductive types, among other features. Even though the base theory is constructive, several common axioms are admissible, including excluded middle. We do not make use of any axioms in this development.

The Mathematical Components libraries, also known as MathComp \cite{MathComp}, are libraries of formalized mathematics originally developed for the mechanization of the Four Color Theorem \cite{Gonthier2008}. They serve as an alternative to Coq's standard library and provide the theories of basic types such as natural numbers and lists (\texttt{mathcomp-ssreflect}), as well as finite sets of so-called choice types (\texttt{mathcomp-finmap}, \cite{finmap}).
  This development is based on MathComp and uses the SSReflect proof language \cite{SSReflect}.

  Other interactive proof assistants could have been used to achieve similar results, but Coq provides many advantages. Its underlying theory is strong enough to prove our results, there are several well-developed libraries for many useful data structures, and the community is large and active. Furthermore, algorithms implemented in Coq can be extracted to other programming languages more suited for computation, such as OCaml. We do not make use of extraction in this development yet, but could do so in the future to obtain a certified and practical decision procedure for $\QRC$.

\subsection{Formalization}
\label{sec:formalization}

This paper tries to be accessible to someone who has never used Coq, or even other interactive proof assistants. For this reason, we mostly highlight the interesting design decisions and difficulties that would plausibly arise in other formalization efforts and stick to standard mathematical notation. The only exception is Section~\ref{sec:coercions}, where we briefly comment on a well-known issue with type hierarchies and the solution we implemented.

When possible, we mention the Coq name for each definition and theorem presented here. These names are hyperlinks to an online rendition of their source code. There is also a summary of the formalization available online,\footnote{\url{https://ana-borges.gitlab.io/QRC1-Coq/v0.1.0/Summary.html}} serving as a kind of documentation.


  In Coq, every term has a type, and every type is also a term (and thus has a (larger) type itself). There is a special type, called \texttt{Prop}, which is meant to represent logical propositions. Thus, when $P : \texttt{Prop}$ we think of $P$ as the statement of a lemma, and of inhabitants of $P$ as proofs of $P$. Most of the time, we don't care which particular proof of $P$ was used to show $P$ was inhabited (i.e., proved).\footnote{The proof mining field \cite{Kohlenbach2008} is a clear exception, although if one were to implement proof mining techniques in Coq, one would probably use something other than \texttt{Prop} to represent logical propositions.} In contrast, when defining a non-\texttt{Prop} object, we often do care about which specific inhabitant was chosen. For example, the statement $0 : \texttt{nat}$ is much more informative than the statement ``\texttt{nat} is inhabited''. We refer to inhabitants of \texttt{Prop} as proofs or non-informative terms, and to other objects as informative terms.

  Even though it is possible, there are some issues with including proofs in the middle of otherwise-informative terms. It has been our experience that a Coq development becomes much simpler when this is avoided and informative terms are clearly separated from non-informative ones.\footnote{This is arguable and boils down to style. There is a well-known Coq textbook \cite{CPDT} describing the opposite strategy. Here we tried to follow the MathComp guidelines \cite{MathCompBook} instead.} We only mix these when defining objects meant exclusively for theorem statements (as in Section~\ref{sec:coercions}), or when we couldn't find an alternative. Even then, postponing this mix as much as possible led to a clear improvement in the complexity of the implementation, as described in Section~\ref{sec:ConstE}.

  We briefly present some figures comparing this formalization with mathematical text describing the same definitions, theorems, and proofs. The formalization described in this document takes up about 8750 Coq words (corresponding to about 1400 lines of code), roughly twice as much as the number of words used to describe the same objects in the \LaTeX{} source for \cite{QRC1} (corresponding to about 8 pages). Strikingly, the ongoing formalization of the completeness theorem is already at almost 19000 Coq words or 3300 lines of code (not counting the code shared with the soundness formalization), while the relevant \LaTeX{} source is about 3500 words or 6 pages long.

\section{Quantified and strictly positive formulas}
\label{sec:language}

We define the names of variables, \coqident{Language}{VarName}, as simply the natural numbers, ensuring that we have a countable number of variables available. We then define the concept of \coqident{Language}{signature} as including a finite set of constant names, a finite set of predicate names, and a function from the predicate names to the natural numbers assigning an arity to each one. Our language includes no non-constant function symbols.

A \coqident{Language}{term} is either a variable or a constant. We define the appropriate canonical instances for \eqType{} (equality on terms is decidable), \countType{} (there is a countable amount of terms) and \choiceType{} (there is a choice operator for terms). This makes it possible to talk about finite sets of terms using the machinery of the Finite Maps Library \cite{finmap} later on.

A \coqident{Language}{formula} is either $\top$, a predicate name together with a tuple of terms of the arity given by the signature, a conjunction of two other formulas, a diamond of one other formula, or a universal quantifier of a variable and another formula. We use the standard mathematical notation in this text, and reasonable approximations for this notation in the Coq development.


The \coqfile{}{Language} file then goes on to define several standard notions and facts about them, such as free variables ($\fv(\varphi)$ or \coqident{Language}{fv}), substitution ($\varphi\subst{t_1}{t_2}$ or \coqident{Language}{sub}), and being free for a variable in a formula (no occurrence of a free variable becomes bound after the substitution, or \coqident{Language}{freefor}), which we discuss in the next subsection.

\subsection{Binders}

Our formulas live in a quantified language, and as such there is a distinction between free and bound variables. This distinction is important when dealing with substitution, since it should not impact bound variables. Thus, $(\Forall{x} \varphi)\subst{x}{y}$ should be exactly $\Forall{x} \varphi$ because $x$ is not a free variable of that formula.

There is one tricky issue, though: cases where replacing a free variable by a term lead to a previously free occurrence becoming bound, such as in $(\Forall{y} S(x, y))\subst{x}{y}$. Here a naive substitution would lead to $\Forall{y} S(y, y)$, which clearly does not preserve logical strength. In informal mathematics it is common to ignore this issue by observing that the names of the bound variables are ultimately irrelevant: if we wish to replace $x$ by $y$ in $\Forall{y} S(x, y)$, then this can be achieved by first renaming the bound variable to some fresh name such as $z$, and then doing the substitution. The final formula would then be $\Forall{z} S(y, z)$. This is the approach taken by O'Connor in his formalization of the Gödel-Rosser incompleteness theorem \cite{OConnor2005}. However, the paper cites this decision as having led to many issues in the formalization; although it ultimately works, we did not wish to use the same strategy.

Another common solution for this problem is to use de Bruijn indexes \cite{deBruijn1972}. This avoids naming bound variables altogether, so this concern does not appear. However, this approach is complex in its own right and would make the formalization considerably different from the original paper.

There is a tool named Autosubst \cite{Stark2019_Autosubst2} that internally uses de Bruijn indexes but generates the boilerplate code by itself and thus cuts back on the complexity and size of the developments. We have not yet made use of Autosubst, but it would be interesting to see how many lines of code and complications it would save. We leave this as future work.

The approach we settled on was inspired by \cite{SernadasSernadas2012} and the will to avoid mixing non-informative and informative objects as explained in Section~\ref{sec:formalization}. We define unguarded substitution, \coqident{Language}{sub}, and add an extra assumption, \coqident{Language}{freefor}, as needed. This assumption assures us that if the substitution goes through then the replacing term will not be captured under any binders.

We already spoke of terms being free for variables in formulas in our previous work \cite{QRC1, Escape}, so the formalization is very similar to the informal mathematics. Furthermore, there were no significant complications in using this approach in the formalization of the Kripke soundness theorem for $\QRC$. This was no longer the case for the formalization of the Kripke completeness theorem, but we postpone discussing this to a future work, when the formalization is completed.

One downside of this strategy is that variable names must be picked with some foresight. Going back to our example from above, if $(\Forall{y} S(x, y))\subst{x}{y}$ ever appears in our development then we can perform the substitution, but won't be able to use any of the results about it because here $y$ is not free for $x$ in $\Forall{y} S(x, y)$. Thus it is assumed that in practice the names for the bound variables do not clash with the names for the free variables, or that bound variables are renamed as needed.



\newpage

\section{\texorpdfstring{$\QRC$}{QRC1}}
\label{sec:QRC1}

The axioms and rules of $\QRC$ are defined in a deeply embedded way in the \coqfile{}{QRC1} file, which also includes some proofs of simple $\QRC$ facts.

\begin{definition}[\coqident{QRC1}{QRC1Proof}]
  Let $\varphi$, $\psi$, and $\chi$ be any quantified strictly positive formulas. The axioms and rules of $\QRC$ are the following:

\begin{multicols}{2}
\begin{enumerate}[label=\upshape(\roman*),ref=\thetheorem.(\roman*)]
  \item $\varphi \leadsto \top$ and $\varphi \leadsto \varphi$;
  \item $\varphi \land \psi \leadsto \varphi$ and $\varphi \land \psi \leadsto \psi$;
  \item if $\varphi \leadsto \psi$ and $\varphi \leadsto \chi$, then\\$\varphi \leadsto \psi \land \chi$;
  \item if $\varphi \leadsto \psi$ and $\psi \leadsto \chi$, then $\varphi \leadsto \chi$;
  \item if $\varphi \leadsto \psi$, then $\Diamond \varphi \leadsto \Diamond \psi$;
  \item $\Diamond \Diamond \varphi \leadsto \Diamond \varphi$;\label{ax:Trans}
  \item if $\varphi \leadsto \psi$, then $\varphi \leadsto \Forall{x} \psi$\\($x \notin \fv(\varphi)$);\label{rule:AllIr}
  \item if $\varphi\subst{x}{t} \leadsto \psi$, then $\Forall{x} \varphi \leadsto \psi$\\($t$ free for $x$ in $\varphi$);\label{rule:AllIl}
  \item if $\varphi \leadsto \psi$, then $\varphi\subst{x}{t} \leadsto \psi\subst{x}{t}$\\($t$ free for $x$ in $\varphi$ and $\psi$);\label{rule:TermI}
  \item if $\varphi\subst{x}{c} \leadsto \psi\subst{x}{c}$, then $\varphi \leadsto \psi$\\($c$ not in $\varphi$ nor $\psi$).\label{rule:ConstE}
\end{enumerate}
\end{multicols}

If $\varphi \leadsto \psi$, we say that $\psi$ follows from $\varphi$ in $\QRC$.
\end{definition}

We briefly comment on the above axioms and rules. The first six statements correspond to axioms and rules of $\RC_1$, while the two quantifier rules are standard in first-order logic. The final two rules, called term instantiation and constant elimination respectively, fulfill an essential role in the completeness of $\QRC$. The best way to think of them is as quantifier rules in disguise. Since our semantics (described in Section~\ref{sec:semantics}) interprets the free variables of both sides of $\leadsto$ in the same way, we can also think of such free variables as being generalized outside this implication. In other words, $P(x) \leadsto Q(x)$ can be thought of as $\Forall{x} (P(x) \leadsto Q(x))$. We never explicitly write the latter, since it falls outside the scope of the strictly positive language. However, we do wish to arrive at the conclusions such a formula promises, namely, we wish to be able to simultaneously instantiate $x$ on both sides of $\leadsto$ by any term (accomplished by Rule~\ref{rule:TermI}) and to simultaneously ``generalize'' a given term as well (accomplished by Rule~\ref{rule:ConstE}).

We used a deep embedding to represent the axioms and rules of $\QRC$, which facilitates the proofs of meta-theorems such as soundness and completeness. However, it is still quite easy to use this embedding to prove theorems of $\QRC$ itself, as the simple formalization of the following lemma illustrates.

\begin{lemma}
\label{lem:QRC1consequences}
  The following are theorems (or derivable rules) of $\QRC$:
  \begin{enumerate}[label=\upshape(\roman*),ref=\thetheorem.(\roman*)]
    \item \coqident{QRC1}{AllC}: $\Forall{x} \Forall{y} \varphi \leadsto \Forall{y} \Forall{x} \varphi$;
    \item \coqident{QRC1}{All_sub}: $\Forall{x} \varphi \leadsto \varphi\subst{x}{t}$ ($t$ free for $x$ in $\varphi$);
    \item \coqident{QRC1}{Diam_All}: $\Diamond \Forall{x} \varphi \leadsto \Forall{x} \Diamond \varphi$;
    \item \coqident{QRC1}{alphaconversion}: $\Forall{x} \varphi \leadsto \Forall{y} \varphi\subst{x}{y}$ ($y$ free for $x$ in $\varphi$ and $y \notin \fv(\varphi)$);
    \item \coqident{QRC1}{TermIr}: if $\varphi \leadsto \psi$, then $\varphi \leadsto \psi\subst{x}{t}$ ($x$ not free in $\varphi$ and $t$ free for $x$ in $\psi$);
    \item \coqident{QRC1}{Const_AllIr}: if $\varphi \leadsto \psi\subst{x}{c}$, then $\varphi \leadsto \Forall{x} \psi$ ($x$ not free in $\varphi$ and $c$ not in $\varphi$ nor $\psi$). \label{item:constants_forall}
  \end{enumerate}
\end{lemma}

Like other provability logics, $\QRC$ is irreflexive, i.e., $\varphi \leadsto \Diamond \varphi$ is not provable. However, unlike other provability logics, this fact can be proved without semantics. Its formalization is called \coqident{QRC1}{Diam_irreflexive}.

\section{Kripke semantics}
\label{sec:semantics}

The Kripke semantics for $\QRC$ generalizes the Kripke semantics for propositional modal logics by transforming each world into a first-order model. Each of the worlds has its own domain, and the only restriction on the domains is that there must be a function between each pair fulfilling certain properties (cf.~Definition~\ref{def:adequate}). We present here the version implemented in Coq and comment on the slight discrepancies with the definition from \cite{Escape} afterward.

  \begin{definition}[\coqident{KripkeSemantics}{rawFrame}, \coqident{KripkeSemantics}{rawModel}]
\label{def:Kripke_model}

  A \emph{Kripke model} $\M$ in a signature $\Sigma$ is a tuple $\la W, R,\allowbreak \{M_w\}_{w \in W},\allowbreak \{\eta_{w, u}\}_{w, u \in W},\allowbreak \{I_w\}_{w \in W},\{J_w\}_{w \in W} \ra$ where:
  \begin{itemize}
    \item $W$ is a finite set (the set of worlds, where individual worlds are referred to as $w, u, v$, etc);
    \item $R$ is a binary relation on $W$ (the accessibility relation);
    \item $M_w$ is a finite set for each $w \in W$ (the domain of the world $w$, whose elements are referred to as $d, d_0, d_1$, etc);
    \item $\eta_{w, u}$ is a function from $M_w$ to $M_u$ for each $w, u \in W$ (the compatibility function between $w$ and $u$);
    \item for each $w \in W$, the interpretation $I_w$ assigns an element of the domain $M_w$ to each constant $c \in \Sigma$, written $c^{I_w}$; and
    \item for each $w \in W$, the interpretation $J_w$ assigns a set of $n$-tuples $S^{J_w} \subseteq \wp((M_w)^n)$ to each $n$-ary relation symbol $S \in \Sigma$.
  \end{itemize}

  The $\la W, R, \{M_w\}_{w \in W}, \{\eta_{w, u}\}_{w, u \in W} \ra$ part of the model is called its \emph{frame}. We say that the frame (or model) is \emph{constant domain} if all the $M_w$ coincide and all the $\eta_{w, u}$ are the identity function.
\end{definition}

The above definition of frame, called \coqident{KripkeSemantics}{rawFrame} because it is not necessarily adequate, is not exactly like the one presented in \cite{Escape}. Note how above we postulate $\eta_{w, u}$ functions for every pair of worlds. In our previous work, $\eta_{w, u}$ was only defined when $w R u$. This made sense because the notion of satisfaction only uses the compatibility functions in those cases. However, including such a non-informative restriction in a Coq definition, although possible, leads to noticeable inconveniences, as described in Section~\ref{sec:formalization}.

Our work-around was to change the notion of frame so that functions $\eta_{w, u}$ must exist for every pair of worlds $w$ and $u$. The fact that in principle we only make use of the ones between pairs of worlds connected through $R$ is immaterial. We must add an extra assumption to the notion of adequate frame to maintain soundness, though: $\eta_{w, w}$ must be the identity for any world $w$.\footnote{This restriction was already implicit for any reflexive world $w$ as a consequence of $\eta_{w, w}$ respecting transitivity in that case (see Definition~\ref{def:adequate}).}
  This decision was crucial in the mechanization of the soundness of Rule~\ref{rule:ConstE}, which was the trickiest one. See Section~\ref{sec:ConstE} for more details.

Note how we do not lose generality with this alternative definition. The extra $\eta_{w, u}$ functions can obviously be dropped to obtain the original definition; on the other hand, as long as the domain $M_u$ is non-empty,\footnote{We never explicitly require non-empty domains, but a $w$-assignment $g$ can only exist if the domain of $w$ is non-empty. Thus, the soundness theorem holds vacuously for empty-domain models, and ignoring such models is not a loss.} we can define a function $\eta_{w, u}$ from $M_w$ to $M_u$. Since this function will not be used, it doesn't matter which one we pick.

The other difference is that we only implement finite models, in the sense that both the set of worlds and each domain are finite. This does not impact the completeness proof, since $\QRC$ has the finite model property \cite{QRC1}, but it does mean that the formalized soundness proof is slightly weaker than the more general one presented in \cite{Escape}.

The relevant frames and models will need to satisfy a number of requisites.

\begin{definition}[\coqident{KripkeSemantics}{adequateF}, \coqident{KripkeSemantics}{adequateM}]
\label{def:adequate}
  A frame $\F$ is \emph{adequate} if:
  \begin{itemize}
    \item $R$ is transitive: if $wRu$ and $uRv$, then $wRv$;
    \item the $\eta$ functions respect transitivity: if $wRu$ and $uRv$, then $\eta_{w, v}(d) = \eta_{u, v}(\eta_{w, u}(d))$ for every $d$ in the domain of $w$; and
    \item the $\eta_{w, w}$ functions are the identity.
  \end{itemize}
  A model is \emph{adequate} if it is based on an adequate frame and it is:
  \begin{itemize}
    \item concordant: if $wRu$, then $c^{I_u} = \eta_{w, u}(c^{I_w})$ for every constant $c$.
  \end{itemize}
  Note that in an adequate and rooted model the interpretation of the constants is fully determined by their interpretation at the root.
\end{definition}

The notion of \coqident{KripkeSemantics}{frame} is defined by pairing a  \coqident{KripkeSemantics}{rawFrame} with a proof that it is \coqident{KripkeSemantics}{adequateF}, and similarly for models. We go into more technical details on Section~\ref{sec:coercions}.

We use assignments to define truth at a world in a first-order model. Fixing a world $w$, a $w$-\coqident{KripkeSemantics}{assignment} $g$ is a function assigning a member of the domain $M_w$ to each variable in the language. 

Two $w$-assignments $g$ and $h$ are \emph{$\Gamma$-alternative}, written $g \xaltern{\Gamma} h$ (or \coqident{KripkeSemantics}{Xaltern}\texttt{ g h }$\Gamma$ in Coq), if they coincide on all variables other than the ones in $\Gamma$. We write $g \xaltern{x} h$ instead of $g \xaltern{\{x\}} h$.
A $w$-assignment $g$ is extended to terms by defining $g(c) := c^{I_w}$ for any constant $c$.

We now define satisfaction at a world.

\begin{definition}[\coqident{KripkeSemantics}{sat}]
\label{def:sat}
  Let $\M = \la W, R, \{M_w\}_{w \in W}, \{\eta_{w, u}\}_{w, u \in W}, \{I_w\}_{w \in W}, \{J_w\}_{w \in W} \ra$ be a model in some signature $\Sigma$, and let $w \in W$ be a world, $g$ be a $w$-assignment, $S$ be an $n$-ary relation symbol, and $\varphi, \psi$ be formulas in the language of $\Sigma$.

  We define $\M, w \Vdash^g \varphi$ ($\varphi$ is true at $w$ under $g$) by induction on $\varphi$ as follows.
  \begin{itemize}

    \item $\M, w \Vdash^g \top$;

    \item $\M, w \Vdash^g S(t_0, \hdots, t_{n-1})$ iff $\la g(t_0), \hdots, g(t_{n-1}) \ra \in S^{J_w}$;

    \item $\M, w \Vdash^g \varphi \land \psi$ iff both $\M, w \Vdash^g \varphi$ and $\M, w \Vdash^g \psi$;

    \item $\M, w \Vdash^g \Diamond \varphi$ iff there is a $u \in W$ such that $w R u$ and $\M, u \Vdash^{\eta_{w, u} \circ g} \varphi$;

    \item $\M, w \Vdash^g \Forall{x} \varphi$ iff for all $w$-assignments $h$ such that $h \xaltern{x} g$, we have $\M, w \Vdash^{h} \varphi$.

  \end{itemize}
\end{definition}

  Note how we haven't required that $\M$ be adequate in the definition of satisfaction, as it is not needed. We will of course assume the models are adequate when proving facts about them.
  Note also that the expression $\M, w \Vdash^g \varphi$ is only defined when $g$ is a $w$-assignment.

The main results on $\QRC$ are as follows.

\begin{theorem}[\coqident{KripkeSemantics}{soundness}]
\label{thm:modal_soundness}
If $\varphi \leadsto \psi$, then for any adequate model $\M$, for any world $w \in W$, and for any $w$-assignment $g$:
  \begin{gather*}
    \M, w \Vdash^g \varphi
    \implies
    \M, w \Vdash^g \psi
    .
  \end{gather*}
\end{theorem}

\begin{theorem}[Completeness, \cite{Escape}]
\label{thm:constant_domain}
  If $\varphi \not\leadsto \psi$, then there is an adequate, finite, constant domain and irreflexive model $\M$, a world $w \in W$, and a $w$-assignment $g$ such that:
  \begin{equation*}
    \M, w \Vdash^g \varphi
    \quad \text{and} \quad
    \M, w \not\Vdash^g \psi
    .
  \end{equation*}
\end{theorem}

Since we have the finite model property, we can conclude that $\QRC$ is decidable by Post's Theorem.

We focus on the (constructive and axiom-free) formalization of the soundness theorem in Section~\ref{sec:soundness} and leave the formalization of the completeness theorem to a future work.

\subsection{Type hierarchies}
\label{sec:coercions}


When defining specific frames or models or operations on arbitrary frames or models such as Definition~\ref{def:sat}, we use the \texttt{raw} versions. On the other hand, when stating facts about frames or models we use the adequate versions, if necessary. We make use of implicit coercions in order to smoothly refer to operations that expect, for example, a \coqident{KripkeSemantics}{rawFrame} in a theorem statement about a \coqident{KripkeSemantics}{frame}.

A coercion is a function $f : A \to B$ that is automatically used by Coq when an otherwise ill-typed statement would be well-typed in the presence of $f$. For example, we declare a coercion from $\coqident{KripkeSemantics}{rawFrame}$ to $\coqident{KripkeSemantics}{world}$ (the set of worlds) that lets us write statements such as \texttt{forall (F : rawFrame), forall (w : F), ...} that closely resemble the common shorthand of stating that a world is part of a frame instead of part of the set of worlds of the frame. In this case Coq automatically infers the implicit coercion \coqident{KripkeSemantics}{world} necessary to make the statement type-check. Explicitly, it would be \texttt{forall (F : rawFrame), forall (w : world F), ...}

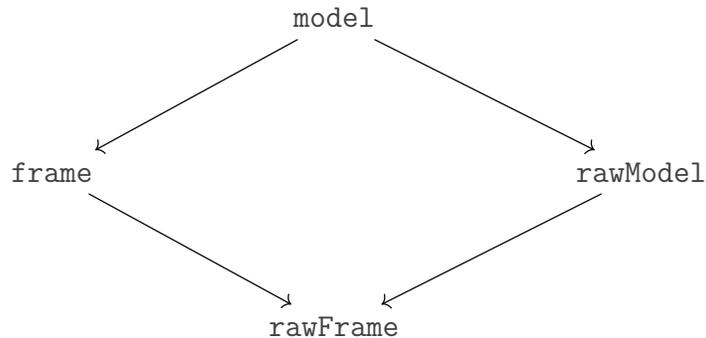
\begin{figure}[ht]
\centering
\begin{tikzcd}
  & & \coqident{KripkeSemantics}{model} \arrow[rrdd] \arrow[lldd] & & \\
  & & & & \\
  \coqident{KripkeSemantics}{frame} \arrow[rrdd] & & & & \coqident{KripkeSemantics}{rawModel} \arrow[lldd] \\
  & & & & \\
  & & \coqident{KripkeSemantics}{rawFrame} & &
\end{tikzcd}
\caption{A representation of the four datatypes defined to represent frames and models and the coercions between them. Each arrow from \texttt{X} to \texttt{Y} represents the coercion \texttt{Y\_of\_X}.}
\label{fig:diamond}
\end{figure}

We use a small number of coercions in our development, the most important of which are represented in Figure~\ref{fig:diamond}. These coercions serve as a translation between a type and a super-type (in the sense that the former is a sub-type of the latter). We have a very small type hierarchy. Formalizations of, say, mathematical algebra or large libraries such as MathComp include rich hierarchies \cite{Sakaguchi2020}, and there are existing tools to implement and maintain such large hierarchies such as the Hierarchy Builder \cite{HierarchyBuilder}.

Still, even with a small hierarchy we do run into some issues.
For example, consider the following unification problem:
\begin{equation}
\label{eq:unification}
  \coqident{KripkeSemantics}{rawFrame_of_frame} \ ?F
    = \coqident{KripkeSemantics}{rawFrame_of_rawModel} \ (\coqident{KripkeSemantics}{rawModel_of_model} \ M)
\end{equation}

In words, given a \coqident{KripkeSemantics}{model} $M$, a \coqident{KripkeSemantics}{frame} $?F$ must be found such that its \coqident{KripkeSemantics}{rawFrame} corresponds to the \coqident{KripkeSemantics}{rawFrame} of the model.
The diamond represented in Figure~\ref{fig:diamond} commutes, and so we define the canonical coercion \coqident{KripkeSemantics}{frame_of_model} as the path to solve \eqref{eq:unification} with $?F = \coqident{KripkeSemantics}{frame_of_model} \ M$.



\section{Soundness}
\label{sec:soundness}

We show the soundness of $\QRC$ (Theorem~\ref{thm:modal_soundness}) by induction on the proof of $\varphi \leadsto \psi$. Some of the axioms and rules are trivial, and we do not comment on them.

The soundness of the transitivity axiom (Axiom~\ref{ax:Trans}, \coqident{QRC1}{Trans}) follows from both the transitivity of $R$ and the fact that the compatibility functions respect transitivity. We also use the fact that assignments only matter for variables appearing free in the formula (Lemma~\ref{lem:sat_Xalternfv}, further discussed in Section~\ref{sec:Xeq}) to take advantage of the extensional equality of $\eta_{w, v}$ and $\eta_{u, v} \circ \eta_{w, u}$.

\begin{lemma}[\coqident{KripkeSemantics}{sat_Xalternfv}]
\label{lem:sat_Xalternfv}
  Let $\M$ be an adequate model, $w$ be any world, $g, h$ be any $\Gamma$-alternative $w$-assignments, and $\varphi$ be a formula with no free variables in $\Gamma$. Then:
  \begin{equation*}
    \M, w \Vdash^g \varphi \iff \M, w \Vdash^{h} \varphi
    .
  \end{equation*}
\end{lemma}

This lemma is all that is needed to show the soundness of the $\forall$-introduction on the right rule (Rule~\ref{rule:AllIr}, \coqident{QRC1}{AllIr}). For both $\forall$-introduction on the left (Rule~\ref{rule:AllIl}, \coqident{QRC1}{AllIl}) and term instantiation (Rule~\ref{rule:TermI}, \coqident{QRC1}{TermI}), we use the fact that a formula $\varphi$ is valid under an assignment $\tilde g$ if and only if $\varphi\subst{x}{t}$ is valid under an assignment $g$ when $g \xaltern{x} \tilde g$ and $\tilde{g}(x) = g(t)$ (Lemma~\ref{lem:substitution_formula}) as the main building block.

\begin{lemma}[\coqident{KripkeSemantics}{substitution_formula}]
\label{lem:substitution_formula}
  Let $\M$ be an adequate model, $w$ be a world, and $g, \tilde{g}$ be $x$-alternative $w$-assignments such that $\tilde{g}(x) = g(t)$. Then for every formula $\varphi$ with $t$ free for $x$:
  \begin{equation*}
    \M, w \Vdash^{\tilde{g}} \varphi
    \iff
    \M, w \Vdash^g \varphi\subst{x}{t}
    .
  \end{equation*}
\end{lemma}
Finally, the soundness of the constant elimination rule (Rule~\ref{rule:ConstE}, \coqident{QRC1}{ConstE}) is the trickiest, and we postpone its discussion to Section~\ref{sec:ConstE}.

\subsection{Finite sets}
\label{sec:Xeq}

We made a decision to only work with finite sets. This allowed us to make use of the nice Finite Maps library for choice types of MathComp \cite{finmap} instead of having to prove many basic facts from scratch. However, Lemma~\ref{lem:sat_Xalternfv} (above) made us momentarily reconsider this decision.

This lemma feels intuitive and in fact its proof was omitted in \cite{QRC1} and \cite{Escape}. However, it is not as straightforward as it looks. A simple induction is underpowered to solve it; one must do induction building with the assumption that $g$ and $h$ are $(\text{Vars} \setminus \fv(\varphi))$-alternative instead. Since $\text{Vars} \setminus \fv(\varphi)$ is not a finite set, it can't be represented by the machinery of the Finite Maps library. In order to get around this, we defined the notion of $\Gamma$-equivalent assignments.

\begin{definition}[\coqident{KripkeSemantics}{Xeq}]
\label{defi:Xeq}
  Two $w$-assignments $g$ and $h$ are said to be $\Gamma$-equivalent if they agree on every variable in $\Gamma$.
\end{definition}

Clearly $g$ and $h$ are $(\text{Vars} \setminus \fv(\varphi))$-alternative if and only if they are $\fv(\varphi)$-equivalent. With this formulation we can prove Lemma~\ref{lem:sat_Xeqfv} by induction first and obtain Lemma~\ref{lem:sat_Xalternfv} as an easy corollary.

\begin{lemma}[\coqident{KripkeSemantics}{sat_Xeqfv}]
\label{lem:sat_Xeqfv}
Let $\M$ be an adequate model, $w$ be a world, $\varphi$ be a formula, and $g, h$ be $\fv(\varphi)$-equivalent $w$-assignments. Then:
  \begin{equation*}
    \M, w \Vdash^g \varphi \iff \M, w \Vdash^{h} \varphi
    .
  \end{equation*}
\end{lemma}

\subsection{Soundness of the constant elimination rule}
\label{sec:ConstE}

Recall the constant elimination rule (Rule~\ref{rule:ConstE}, \coqident{QRC1}{ConstE}):
\begin{center}
  if $\varphi\subst{x}{c} \leadsto \psi\subst{x}{c}$, then $\varphi \leadsto \psi$\\($c$ not in $\varphi$ nor $\psi$)
\end{center}

The argument for its soundness goes as follows. Suppose that $\varphi\subst{x}{c} \leadsto \psi\subst{x}{c}$ is sound and that $\M, w \Vdash^g \varphi$ for some adequate model $\M$, world $w$, and $w$-assignment $g$. We wish to show that $\M, w \Vdash^g \psi$. We build a new model $\M\rsubst{w}{c}{g(x)}$ that is identical to $\M$ except it interprets $c$ as $g(x)$ in $w$, in hopes that $\M\rsubst{w}{c}{g(x)}$ satisfies $\chi\subst{x}{c}$ if and only if $\M$ satisfies $\chi$, for any formula $\chi$ where $c$ does not appear. We can then deduce that $\M\rsubst{w}{c}{g(x)}, w \Vdash \varphi\subst{x}{c}$ from our assumption that $\M, w \Vdash^g \varphi$, and, since $\varphi\subst{x}{c} \leadsto \psi\subst{x}{c}$ is sound, this means that $\M\rsubst{w}{c}{g(x)}, w \Vdash^g \psi\subst{x}{c}$, and consequently that $\M, w \Vdash^g \psi$.

The above proof sketch should be intuitive enough, but it omits a crucial point: the naive definition of $\M\rsubst{w}{c}{g(x)}$ is not concordant, because the interpretation of a constant is being changed at $w$ without being changed anywhere else. It is fine to propagate the change to the successors of $w$ through the compatibility functions, and this would restore the concordance if $w$ were the root of the model. However, when $w$ is not the root, there is no clear solution other than dropping every other world from the model, which is what is done in \cite{QRC1}. It works well because the satisfaction of a formula at $w$ depends only on the model restricted to $w$ and its successors.

We originally tried to implement this proof directly: restrict $\M$ to $w$ and its successors and then replace the interpretation of $c$ with $g(x)$ at $w$ and with $\eta_{w, u}(g(x))$ at all the successors $u$ of $w$. In this proof, the models are adequate every step of the way. However, implementing this strategy proved rather difficult. A model restricted to $w$ and its successors is naturally defined as a regular model together with a non-informative statement to the effect that every world is either $w$ or its successor. Then the next step would be to define a way to change the interpretation of a constant at the root and propagate it to all its successors. However, trying to do this on top of restricted models proved hard, in part because there is no built-in concept of root. Adequate models do not need to be rooted and we didn't want to include this restriction.

Instead, we ended up changing the proof to postpone including non-informative elements as much as possible. The key insight is that only the final model needs to be adequate, and so we can change the constant interpretation first and only then restrict the worlds to obtain concordance. Here is also where the decision to have compatibility functions for every pair of worlds shines, as we'll soon see. We define the constant interpretation for the new model as follows.

\begin{definition}[\coqident{KripkeSemantics}{replace_I}]
\label{def:replace_I}
Let $\M$ be a model, $w$ be a world, $c$ be a constant, and $d$ be an element of the domain of $w$. If $I$ is the constant interpretation of $\M$, we define a new interpretation $I\rsubst{w}{c}{d}$ as follows. For a given world $u$, $c^{I\rsubst{w}{c}{d}_u} := \eta_{w, u}(d)$. $I\rsubst{w}{c}{d}$ behaves like $I$ for every other constant.
\end{definition}

Note that the above definition is well-typed even if $\M$ is not an adequate model, and it won't lead to an adequate model unless $w$ happens to be the root of $\M$. Note also that, if $\M$ is adequate, then $c^{I\rsubst{w}{c}{d}_w} = \eta_{w,w}(d) = d$, because $\eta_{w, w}$ is the identity in adequate models.

Finally, observe that if $\eta_{w, u}$ only existed when $w R u$, we could not have defined $I\rsubst{w}{c}{d}$ like this, for there would be no way to obtain an element of the domain of $u$ in the cases where $u$ was not a successor of $w$. Recall that we're going to drop these worlds later anyway, so it doesn't matter which domain element this is; only that we have one in hand. Even though this could have been implemented in other ways (for example, by designating a default element for each domain), this particular solution is elegant in its simplicity and symmetry, as there is no need to have a case distinction on $w R u$.

The first approximation to $\M\rsubst{w}{c}{g(x)}$ is then a copy of $\M$ with the constant interpretation replaced by $I\rsubst{w}{c}{g(x)}$. We can already prove the desired property about this model (\coqident{KripkeSemantics}{sat_replace}), namely that $\M\rsubst{w}{c}{g(x)}$ satisfies $\chi\subst{x}{c}$ at $w$ if and only if $\M$ satisfies $\chi$ at $w$, for any formula $\chi$ where $c$ does not appear. It now remains to further modify $\M\rsubst{w}{c}{g(x)}$ so that it is adequate, by dropping all spurious worlds, obtaining $\M\rsubst{w}{c}{g(x)}\rest{w}$. The final model, called \coqident{KripkeSemantics}{restrict_replace}, is finally adequate and allows us to prove the desired result, Lemma~\ref{lem:sat_restrict_replace}, which has the soundness of the constant elimination rule as a corollary.

\begin{lemma}[\coqident{KripkeSemantics}{sat_restrict_replace}]
\label{lem:sat_restrict_replace}
Given a constant $c$, a formula $\varphi$ where $c$ does not appear, an adequate model $\M$, a world $w$, and a $w$-assignment $g$, we have:
  \begin{equation*}
    \M, w \Vdash^{g} \varphi
    \iff
    \M\rsubst{w}{c}{g(x)}\rest{w}, w \Vdash^{g} \varphi\subst{x}{c}
    .
  \end{equation*}
\end{lemma}

\section{Conclusions and future work}
\label{sec:concl}

In this work we presented a Coq mechanization of the $\QRC$ logic, its Kripke semantics, and a formalized proof of its soundness theorem. We discussed the difficulties in translating these objects and results to Coq as well as our proposed solutions. The formalization process suggested a slightly different definition of Kripke model that is nevertheless equivalent to the previous one under common assumptions. This new definition allowed for a simpler soundness proof.

The clear next step is to formalize Theorem~\ref{thm:constant_domain}, the completeness of $\QRC$, possibly making use of Autosubst \cite{Stark2019_Autosubst2} to ease complications with binders. With both the axiom system and the completeness proof, it should be possible to generate a mechanized and formalized decision procedure for $\QRC$ via Post's Theorem, which has already been formalized itself \cite{CoqUndecidability}. It would also be interesting to see if some of the techniques described in the recent formalization of a decision procedure for $\GL$ in HOL Light \cite{MaggesiBrogi2022} are applicable, since $\GL$ is a closely related logic.

Other modal results on $\QRC$ could be formalized too, such as the fact that it is the strictly positive fragment of the quantified modal logics between $\mathsf{QK}4$ and $\mathsf{QGL}$ \cite{Escape}. Finally, the arithmetical results could be an interesting subject, although these are less elementary and would need to be part of a larger project including practical definitions of arithmetical theories such as Peano Arithmetic and its fragments. The Undecidability Library \cite{CoqUndecidability} might be a good basis for such a project.

\paragraph{Acknowledgments}
  The author wishes to thank the Coq community for its ready support during the formalization process, Joost J. Joosten for proposing this article could be written, Mireia González Bedmar for her comments on an early draft, and the anonymous reviewers for their helpful suggestions.


\newcommand{\noopsort}[1]{}

\end{document}